\documentclass[twocolumn,amssymb,prl,aps]{revtex4-1}
\usepackage{amsmath}
\usepackage{amssymb}
\usepackage{graphicx}
\usepackage{color}

\begin{document}

\title{Spontaneous ordering of magnetic particles in liquid crystals: From chains
to biaxial lamellae}
\author{Stavros~D.~Peroukidis}
\author{Sabine~H.L.~Klapp}
\affiliation{Institute of theoretical Physics, Secr. EW 7-1, Technical University of Berlin, Hardenbergstr. 36, D-10623 Berlin, Germany}
\date{March 27, 2015}

\begin{abstract}
Using Monte Carlo (MC) computer simulations we explore the self-assembly and ordering behavior of a hybrid, soft magnetic system consisting 
of small magnetic nano-spheres in a liquid-crystalline (LC) matrix. Inspired by recent experiments with colloidal rod-matrices we focus
on conditions where the sphere and rod diameters are comparable. Already in the {\em absence} of a magnetic field, 
the nematic ordering of the LC can stabilize formation of magnetic chains along the nematic or smectic director, yielding a state with local (yet no macroscopic) magnetic order. The chains, in turn, increase the overall nematic order, reflecting the complex interplay
of the structure formation of the two components. Increasing the sphere diameter the spontaneous uniaxial ordering is replaced by {\em biaxial} lamellar morphologies 
characterized by alternating layers of rods and magnetic chains oriented perpendicular to the rod's director.
These novel ordering scenarios at zero field suggest
complex response of the resulting hybrid to external stimuli such as magnetic fields and shear forces.
\end{abstract}
\maketitle

%
Mixtures of magnetic nanoparticles (MNP) in liquid-crystalline (LC) matrices attract attention in fundamental physics and material science since about four decades ago: 
in 1970, Brochard and de Gennes \cite{Brochard1970} predicted, based on free energy considerations,
various new effects including spontaneous magnetization in the liquid state, "compensated" phases and giant field-induced effects. Whereas the field sensitivity
has been investigated and verified in many experiments \cite{Liebert1979,Chen1983,Neto1986,Podoliak2011,Kopacansky2008,Buluy2011,Kredentser2013}, new interest was stimulated by the recent discovery 
of spontaneous magnetic ordering in a hybrid system of large, micron-sized magnetic plates embedded
in a lyotropic nematic LC \cite{Mertelj2013}. A crucial ingredient for the observed behavior are the LC-mediated interactions
stemming from local distortions of the LC director via the presence of the plates, whose size is much larger
than that of the LC molecules. Indeed, LC-mediated elastic interactions and their consequences for colloidal self-assembly and ordering have
intensively studied by experiments \cite{Lapointe2009} and simulations \cite{Ravnik2007,Lint2014} for many (non-magnetic) colloid/LC mixtures. 

In the present letter we focus on the opposite case of small MNPs in LC matrices, where the sizes are comparable and distortions of the LC do not play a significant role. An example of such systems are lyotropic suspensions of colloid pigment rods  and magnetite MNPs  which have been recently studied to investigate field-induced ordering effects \cite{May2014,privcomm}. However, although there is strong interest in using such hybrids as new, stimuli-responsive or even adaptive materials \cite{Mertelj2013},
the microscopic structure of small MNPs in LC matrices is essentially unexplored; although this is clearly a prerequisite in terms of advancing new functionalities.
Experimentally, the challenge is to actually "see" the relatively small MNPs with typical
diameters of 10-15 nm; a solution of this problem might be provided by recently developed tomography techniques \cite{Odenbach2014}. Suprisingly, however, there is also
a lack of computer simulation studies (contrary to the manifold of studies on non-magnetic rod-sphere mixtures\cite{Antypov2004,Dogic2000,Peroukidis2010}); indeed, to our knowledge ferronematics have so far only been
studied on the basis of Landa-de Gennes free energy approaches \cite{Zadorozhnii2006,Raikher2013,Podoliak2011}. Thus, microscopically, even basic questions are currently open, such as:
do we find any self-assembly or collective ordering of the MNPs, how do they influence the LC and vice versa, and what is the role of the rod-sphere size ratio? 

To explore these questions we here employ canonical MC simulations \citep{AllenBook} of a simple
binary model system consisting of
$N_{\text{r}}$ rods, represented as prolate, non-magnetic ellipsoids and $N_{\text{s}}$ MNPs, represented as dipolar soft spheres (DSS), 
i.e., spheres with an embedded point dipole moment $\boldsymbol{\mu}_{i}$
($i=1,\ldots,N_{\text{s}}$) in their center.
The rods interact via the anisotropic Gay-Berne (GB) potential involving both, repulsive and attractive (van-der-Waals) interactions, and we use a standard parameterization \cite{Antypov2004}. The length ($l$)-to-
width ($\sigma$) ratio is set to $l/\sigma=3$; this value is well-studied for pure GB systems \cite{Miguel91}, and it is also in the experimentally accessible range for LC/MNP mixtures \cite{privcomm}. The dipolar spheres
have diameter $\sigma_{\text{s}}$. We focus on systems with {\it comparable} sizes with 
$\sigma_{\text{s}}^{*}=\sigma_{\text{s}}/\sigma=1$ and $\sigma_{\text{s}}^{*}=2$. 
The spheres interact via a combination of a soft-sphere potential and the conventional dipole-dipole potential. The reduced dipole moment is set
to $\mu^*=\mu/ \sqrt{\epsilon_0\sigma_{\text{s}}^3}=3$ and we consider reduced temperatures $T^*=k_{\text{B}}T/\epsilon_0 $ (where $k_{\text{B}} $  is the Boltzmann constant and $\epsilon_0$ is
the soft-sphere energy parameter) in the range
$0.6$ to $2.5$. Thus, the coupling parameter $\lambda= \mu^2/\textit{k}_{B}\textit{T}\sigma_{\text{s}}^3 $ takes values between $\approx 3.5$ and $\approx 15$, consistent with values characterizing
real, strongly coupled MNPs \cite{Klokkenburg06,Cobalt2011,Sreekumari2013}. Long-range interactions are treated with the three-dimensional
Ewald sum \cite{Ewald07}. Finally, the interactions between rods and spheres are described
via a modified GB potential \cite{Antypov2004,Cleaver96}, parametrized such that the spheres favor the sides of the rods. 
We have examined systems for a variety of total densities $\rho^{*}=N\sigma^3/V$, compositions $x_{\text{a}}=N_{\text{a}}/N$ (where a=r, s), and resulting 
partial volume fractions $\phi_{\text{a}}=N_{\text{a}}v_{\text{a}}/V$ (where $v_a$ is particles' and $V$ the total volume).
For all chosen systems, we have performed cooling and heating series by varying the temperature
at constant $\rho$ and $x_{\text{a}}$. 

To quantify the degree of overall orientational ordering we have calculated the nematic order parameters of each species, $S^{(\text{a})}$, 
by diagonalizing the corresponding ordering matrix \cite{Camp1999} and extracting the largest eigenvalues. The corresponding eigenvectors yield
the directors $\boldsymbol{\hat{n}}_{\text{s}}$ and $\boldsymbol{\hat{n}}_{\text{r}}$ of the DSS and the rods, respectively. From these we calculate the degree of magnetization (if present),
$P_{1}=\left\langle |\sum_{i=1}^{N_{{\text{s}}}}\boldsymbol{\hat{\mu}}_{i}\cdot\boldsymbol{\hat{n}}_{{\text{s}}}|/N_{\text{s}}\right\rangle $, and 
the biaxiality parameter $B=\left\langle\left[ 3\left(\boldsymbol{\hat{n}}_{\text{r}}\cdot\boldsymbol{\hat{n}}_{\text{s}} \right)^{2}-1\right]/2\right\rangle$  \cite{Cuetos2008}. Finally, the translational structure is analyzed by various pair correlation functions, see \cite{Supplem} for details.

We start by considering the case $\sigma_{\text{s}}^*=1$ and $x_{\text{r}}=0.8$. This corresponds to a ratio of volume fractions of
$\phi_{\text{r}}/\phi_{\text{s}}\cong 12 $, which is in the experimentally accessible range \cite{privcomm}.
As a reference for our GB/DSS simulations we have first considered mixtures of GB rods and unpolar soft spheres.
At the thermodynamic conditions considered, these display fully miscible isotropic (I), nematic (N) and smectic-B (SmB) phases without any indication of global demixing 
or microphase separation \cite{paperext}(contrary to, e.g., mixtures of GB rods and attractive Lennard-Jones spheres \citep{Antypov2004}). Indeed, the miscibility of our reference system is a non-trivial finding
given the well-established fact that the amount of nanoparticles that can be dispersed in LC phases crucially depends on the spheres' chemical affinity \cite{Peroukidis2010}. 

\begin{center}
  \begin{figure}
  \includegraphics[scale=0.43,natwidth=1529,natheight=566]{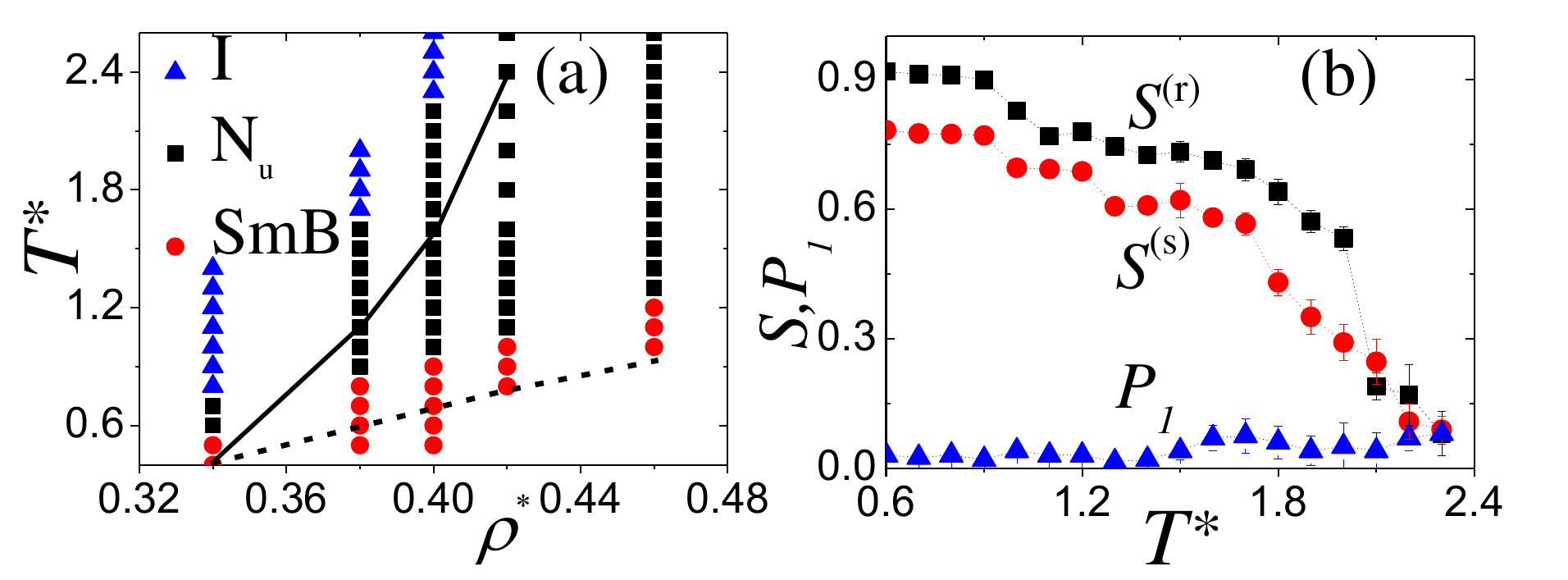}
  \caption{(Color online) (a) State diagram of a mixture of GB rods and dipolar spheres with $\sigma_{\text{s}}^*=1$ ($x_{\text{r}}=0.8$), involving
fully miscible isotropic (I), uniaxial nematic (N$_{\text{u}}$) and uniaxial smectic-B (SmB) states. 
 The symbols indicate state points where the actual simulations were performed. The solid and dashed lines
 indicate state transformations of the non-magnetic reference system. (b) Global order parameters as functions of $T^{*}$
 at $\rho^*=0.40$.}
  \label{fig:01}
  \end{figure}
\end{center}

We now turn to magnetic mixtures with $\sigma_{\text{s}}^*=1$ and $x_{\text{r}}=0.8$. In Fig.~\ref{fig:01}(a) we present a corresponding state diagram
in the $T^{*}$-$\rho^{*}$ plane, obtained by MC simulations with $N=N_{\text{r}}+N_{\text{s}}=720$ particles (test simulations with $N=2000$ and $N=4000$ particles revealed no appreciable differences). We note that Fig.~\ref{fig:01}(a) gives an overview of the nature of equilibrated states rather than true phase transition lines; obtaining the latter requires extensive 
free energy energy calculations not performed here.

From the perspective of the LC host matrix, the overall state behavior of the system is similar to that of the non-magnetic reference mixture. 
An example of the rod's overall ordering behavior at fixed density is given by the plot of the function
$S^{(\text{r})}(T^{*})$ in Fig.~\ref{fig:01}(b). On cooling, $S^{(\text{r})}$ shows a step-like increase from values essentially zero (I state) to $\approx 0.9$ in the SmB state.
We have verified the presence of SmB ordering via the pair correlation function along the rod's director, $g_{\Vert;\boldsymbol{\hat{n}}_{\text{r}}}^{\left({\text{r}}\right)}\left(r_{\Vert}\right)$,
which displays clear differences between the nematic, translationally disordered state and the smB state characterized by formation of  layers with local hexagonal-like order (see \cite{Supplem}). Quantitatively,
we see from Fig.~\ref{fig:01}(a) that the temperatures related to state boundaries are generally larger than in the non-magnetic case, with the gap between the systems increasing with increasing density. In other words,
the stability of orientationally ordered phases is {\em enhanced}, despite of the absence of an external magnetic field.

\begin{center}
  \begin{figure}
  \includegraphics[scale=0.38,natwidth=1766,natheight=1185]{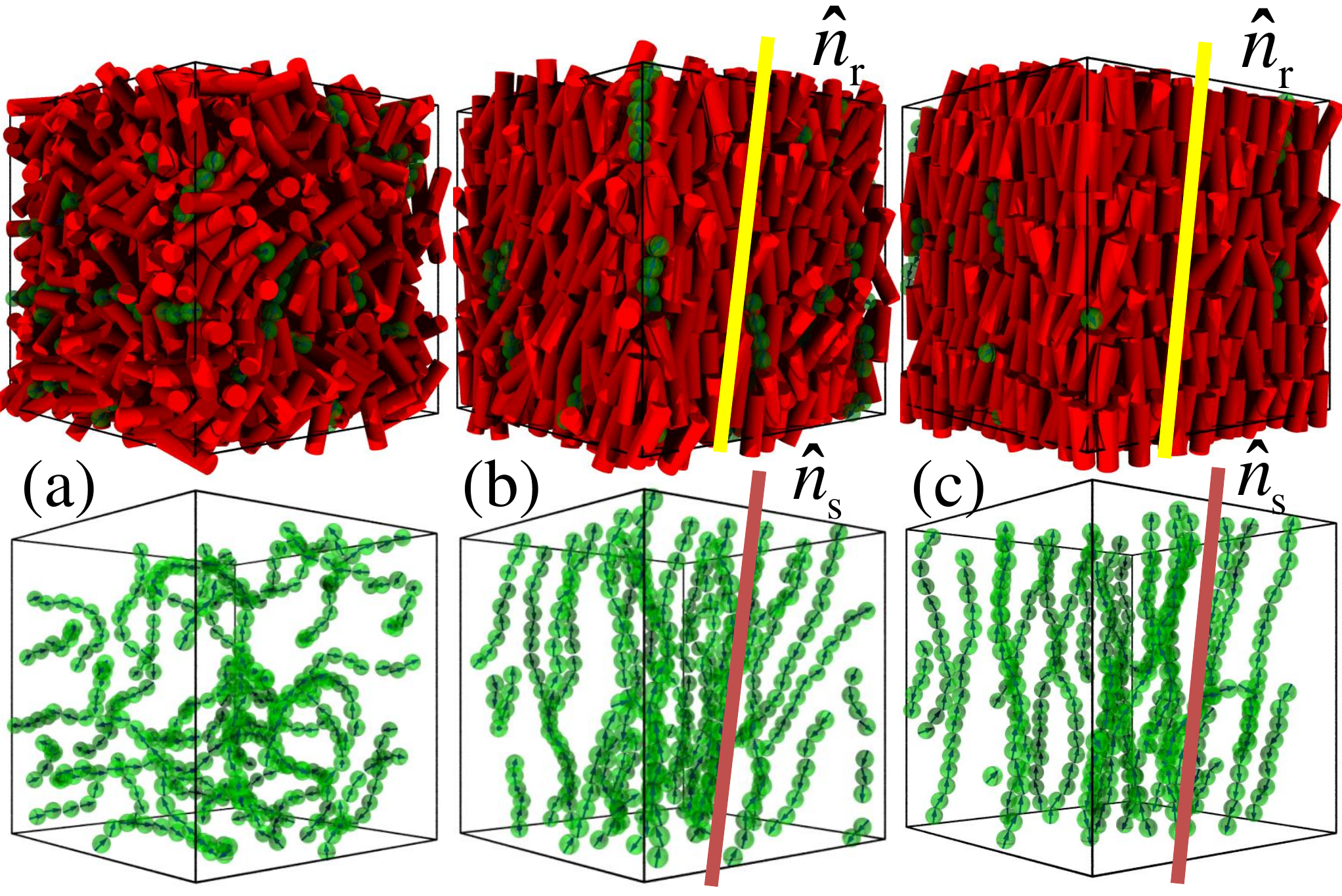}
  \caption{(Color online) Representative simulation snapshots for mixtures with $\sigma_{\text{s}}^*=1$ and $x_{\text{r}}=0.8$
  in a) the isotropic (I) state [$(T^*,\rho^{*})=(1.1,0.34)$], b)
 uniaxial nematic (N$_{\text{u}}$) state [$(T^*,\rho^{*})=(1.2,0.40)$], c) uniaxial smectic-B (SmB) state [$(T^*,\rho^{*})=(0.9,0.40)$]. In the bottom parts
 only the MNPs are shown for clarity. The directors $\boldsymbol{\hat{n}}_{\text{r}}$ and $\boldsymbol{\hat{n}}_{\text{s}}$ are indicated
 by thick lines.}
  \label{fig:02}
  \end{figure}
\end{center}

To understand the underlying mechanisms we consider the microscopic structure of the MNPs in the various LC states illustrated in Fig.~\ref{fig:02}. 
Within the isotropic range [Fig.~\ref{fig:02}(a)] 
the dipolar spheres tend to self-assemble into worm-like chain structures, yet without net alignment [i.e., $S^{(\text{s})}\approx 0$, see also Fig.~\ref{fig:01}(b)]. This behavior
is familiar from pure, strongly coupled DSS systems at low densities \cite{Klapp05}  (recall that the temperature range
in Fig.~\ref{fig:01} corresponds to large dipolar coupling strengths of $\lambda\gtrsim 3.5$). The strongly correlated nature of these chains is also reflected
by periodic arcs of the correlation function $g^{\left({\text{s}}\right)}\left(r_{\Vert},r_{\bot}\right)$ along the dipole moment of a particle, i.e., as function of $r_{\parallel}$
(see \citep{Supplem}). Upon entering the nematic state, the chains spontaneously "unwrap", yielding essentially straight chains 
with head-to-tail alignment of the dipole moments. Moreover, the axes of these magnetic chains form a director $\boldsymbol{\hat{n}}_{\text{s}}$ which coincides (on the average) with the LC director $\boldsymbol{\hat{n}}_{\text{r}}$. This 
yields an {\em uniaxial nematic} (N$_{\text{u}}$) ordering of the hybrid system ($B\approx 1$). 
In fact, the magnetic chains stabilize the LC ordering, as indicated by the enhancement of I-N$_u$ transition temperatures in Fig.~\ref{fig:01}(a). Vice versa, 
the LC stabilizes the mutual alignment of the chains, which is absent in pure DSS fluids at the densities considered here \cite{Klapp05,Jordan2011}.
We note, however, that while the chains themselves form ferromagnetic entities, the overal
magnetization is zero, as reflected by the negligible values of $P_1$ plotted in Fig.~\ref{fig:01}(b). 
Indeed, the absence of 
magnetic correlations beyond a single chain is also reflected by the function $g_{1;\boldsymbol{\hat{n}}_{\text{s}}}\left(r_{\bot}\right)$, which takes large positive values near the origin, but 
then quickly decays to zero (see \cite{Supplem}). Further, there are no indications for a mutual ordering of the chains in the plane perpendicular to $\boldsymbol{\hat{n}}_{\text{r}}$.
Rather, the chains form a random "up-down" structure [see Fig.~\ref{fig:02}(b)].
The ordering {\em within} the ferromagnetic chains becomes even more pronounced upon further cooling the hybrid into the SmB state of the LC host.
This is directly visible by the snapshots in Fig.~\ref{fig:02}(c) and also reflected by the larger values of the order parameter $S^{(\text{s})}$ (as compared to the N$_u$ state).
On the other hand, the translational structure of the chains lateral to the director
remains disordered, contrary to the local hexagonal ordering of the rods within the smectic layers \cite{Supplem}. We term these novel complex fluids uniaxial ferrosmectics.

The fact that oriented chains of MNPs inside a nematic LC stabilizes the host's ordering has been found in several experiments \citep{Chen1983,Neto1986,Kopacansky2008,Podoliak2011}, and has also been theoretically analyzed
by minimizing the elastic free energy \citep{Kredentser2013}. The essential idea is that the presence of long magnetic chains enhances the strength of the local director field of the rods. However,
in all these cases the magnetic ordering was induced by an {\em external} magnetic field. In our case, the self-assembly of the MNPs
occurs spontaneously as a result of the strong dipole-dipole interactions, yielding worm-like chains
already in the isotropic regime. As soon as these chains have been aligned by the LC host, 
the elastic distortions onto the LC host are similar to those in an external field, explaining the LC's enhanced stability.

Not suprinsingly, the effects reported here are sensitive to the size ratio $\sigma_{\text{s}}^{*}$ of the spheres relative to the GB rod's width. Indeed, MC simulations (to be reported elsewhere \cite{paperext})
of systems with significantly smaller dipolar spheres ($\sigma_{\text{s}}^*=0.25$) show that the 
sphere's structure is characterized by rings, chains, and branched structures in {\em all} LC phases. The nematic ordering of the LC 
then leads to an overall alignment of these highly correlated structures. Mixtures
with somewhat larger spheres such as $\sigma_{\text{s}}^*=1.5$ behave similar to the case $\sigma_{\text{s}}^*=1$; 
however, at $\sigma_{\text{s}}^*=2$ the picture changes dramatically. 

\begin{center}
  \begin{figure}
\includegraphics[scale=0.42,natwidth=1538,natheight=575]{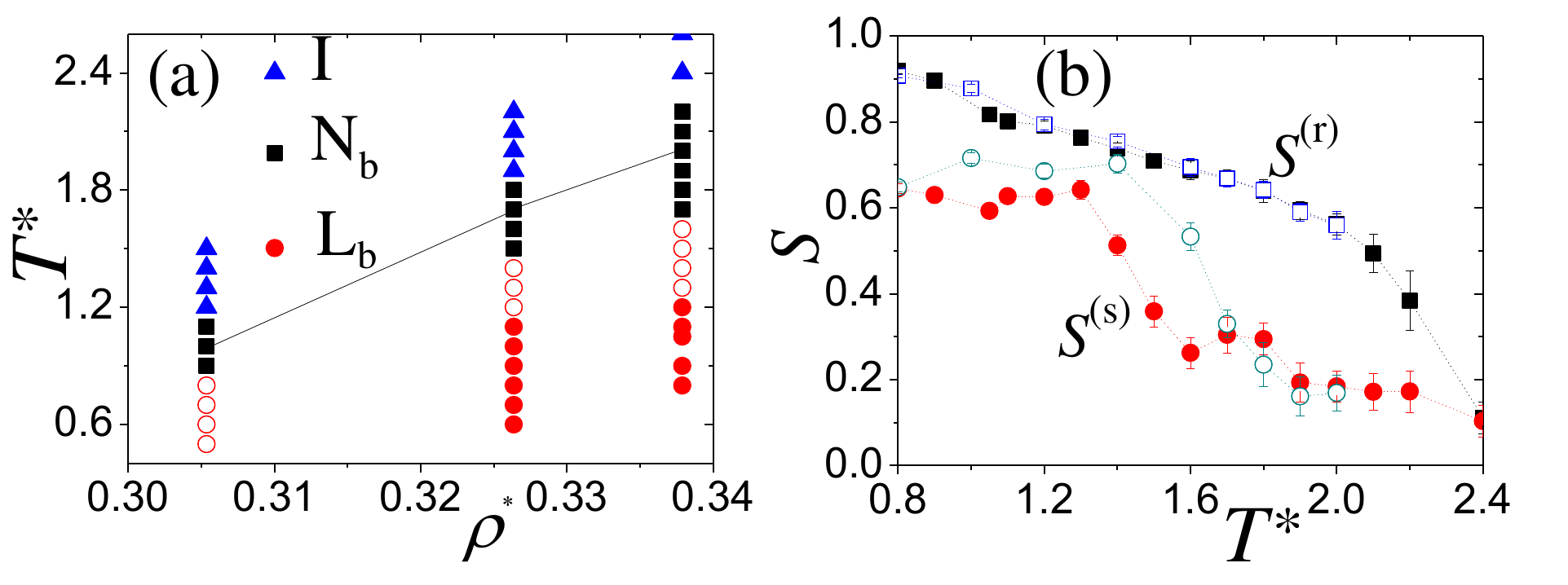}
  \caption{(Color online) (a) State diagram at size ratio $\sigma_{\text{s}}^*=2$ ($x_r=0.9$), involving
fully miscible isotropic (I), biaxial nematic (N$_{\text{b}}$) and biaxial lamellar (L$_{\text{b}}$) states. The solid line
 indicates the I-N transformation of the non-magnetic reference system.
 Open symbols indicate defect-rich states ($\rho^*=0.305$) or hysteresis
 in cooling-heating series. (b) Functions $S^{(a)}(T^{*})$ at $\rho^*=0.338$. Cooling: solid symbols; heating: open symbols. }
  \label{fig:03}
  \end{figure}
\end{center}

As an example we consider systems at $x_{\text{r}}=0.9$. This corresponds to a volume fraction ratio of $\phi_{\text{r}}/\phi_{\text{s}}\cong 3.4$, which is at the lower limit achievable in real systems \cite{privcomm}. Figures~\ref{fig:03}(a)-(b) give an overview of the possible states (obtained with $N=1251$) and the temperature dependence of the order parameters, respectively. 
The I state is structurally similar to the case  $\sigma_{\text{s}}^{*}=1$ in that the 
MNPs form worm-like chains, but no global alignment.
At low temperatures, both the LC and the MNPs exhibit global orientational order, with the characteristics, however, being markedly different from the case $\sigma_{\text{s}}^{*}=1$. First, the boundary between isotropic and orientationally ordered states
is only slightly enhanced relative to the non-magnetic reference case. This already indicates that the MNPs exert a stronger perturbation on the LC host. The second, more striking difference is 
that the low-temperature phases at $\sigma_{\text{s}}^*=2$ are no longer uniaxial, but {\em biaxial} in the sense that the director of the magnetic chains stands {\em perpendicular} to that of the GB rods.

\begin{center}
  \begin{figure}[h!]
  \includegraphics[scale=0.73,natwidth=985,natheight=738]{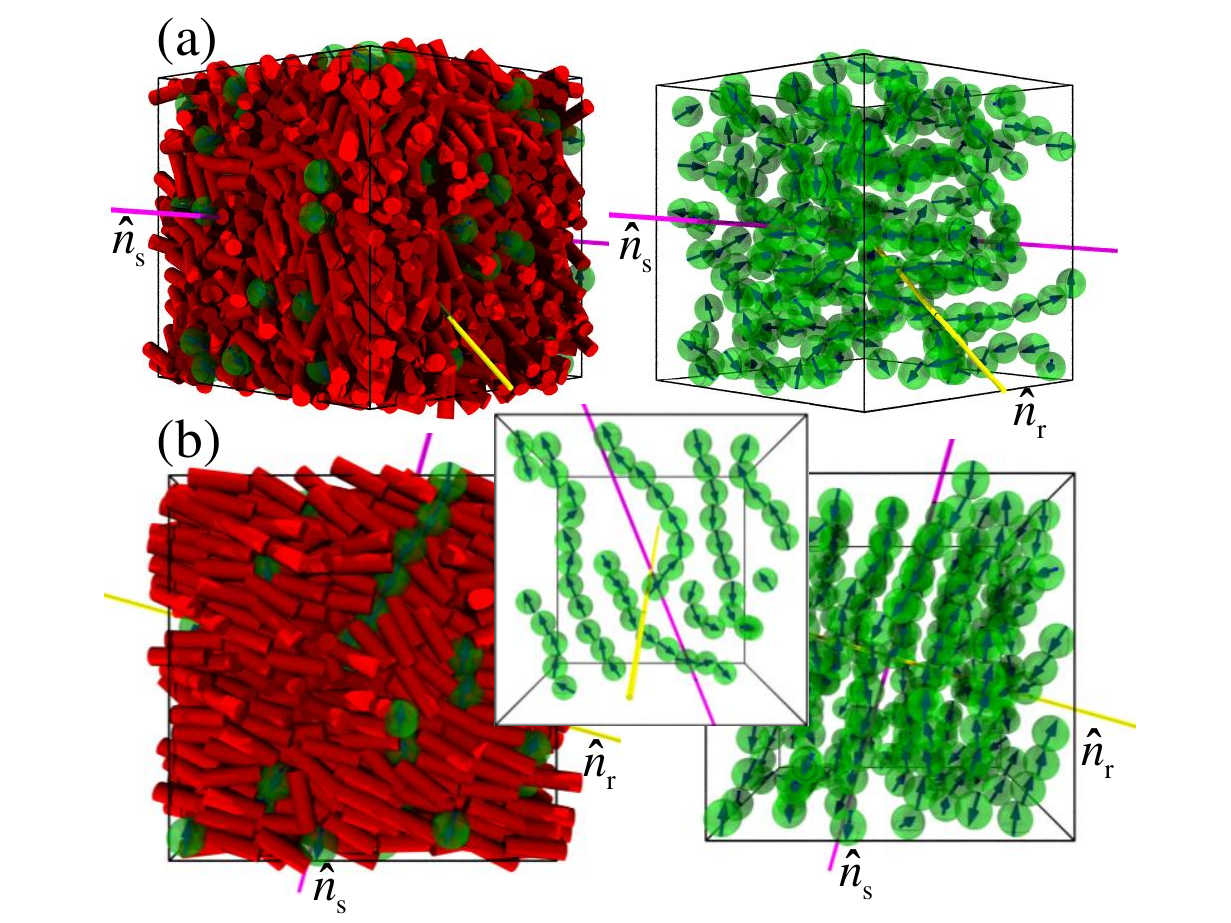}
  \caption{
 (Color online) Representative simulation snapshots for mixtures with $\sigma_{\text{s}}^*=2$ and $x_{\text{r}}=0.9$
 in a) the biaxial nematic (N$_{\text{b}}$) state [$(T^*,\rho^{*})=(1.8,0.338)$] and b) the biaxial lamellar (L$_{\text{b}}$) state [$(T^*,\rho^{*})=(1.2,0.338)$].
 On the right sides \, only the MNPs are shown for clarity. 
 Part b) additionally includes a snapshot of the structure in one dipolar layer.
 The directors $\boldsymbol{\hat{n}}_{\text{r}}$ and $\boldsymbol{\hat{n}}_{\text{s}}$ are indicated
 by thick lines.}
  \label{fig:04}
  \end{figure}
\end{center}

We first consider the biaxial nematic (N$_{\text{b}}$) state. As seen from Figs.~\ref{fig:03}(b) and 4(a), this state is characterized by pronounced nematic ordering of the rods
($0.4\lesssim S^{({\text{r}})}\lesssim 0.7$) without any positional ordering along the (rod) director. Within this ordered matrix,
the DSS arrange into chains which tend to align, yet not as pronounced ($0.15\lesssim S^{({\text{r}})}\lesssim 0.3$)
as in the case $\sigma_{\text{s}}^{*}=1$. The loose ordering of the chains is also visible from the
snapshot presented in Fig.~\ref{fig:04}. The most striking finding, however, is that the principal directors of the magnetic chains and the rods are, on average, {\em perpendicular} to each other
(i.e., $B\cong-0.5$). 
To check for finite-size effects we have performed simulations at $N=2536$ with no appreciable differences; 
the directors always arrange spontaneously in an orthogonal way. We conclude that the system is in a
{\em biaxial} nematic state, contrary  to mixtures with size ratios $\sigma_{\text{s}}^*\lesssim 1.5$ that exhibit uniaxial nematic states.

A further new feature not occuring at smaller values of $\sigma_{s}^{*}$ concerns the low-temperature behavior. As seen from snapshots in Fig.~\ref{fig:04}, 
the system forms a biaxial, {\em lamellar}-like (L$_{\text{b}}$) state characterized by microphase separation into alternating, rod-rich and sphere-rich regions, with a considerable degree of interdigitation. 
In detail, the rods are arranged in smectic-like layers perpendicular $\boldsymbol{\hat{n}}_{\text{r}}$, whereas the dipolar spheres form aligned chains in thinner layers arranged 
{\em in between} the smectic layers of rods. 
These chains point randomly parallel or antiparallel along the sphere director $\boldsymbol{\hat{n}}_{\text{s}}\perp\boldsymbol{\hat{n}}_{\text{r}}$, with the overall degree of 
alignment being substantially larger ($S^{(\text{s})}\approx 0.6$) than in the nematic phase. The net magnetization is zero. Interestingly we have found,
by analyzing appropriate correlation functions, that the {\it local} structure of chains within a layer and with neighboring layers somewhat resembles a two-dimensional rhombic lattice
in the plane perpendicular to the chains's director \cite{Supplem}. Also, we have confirmed the 
 reversibility of the cooling-heating procedure by examining the pair correlations \cite{Supplem}; for example,
 both $g_{\Vert;\boldsymbol{\hat{n}}_{\text{r}}}^{\left({\text{r}}\right)}\left(r_{\Vert}\right)$ and $g_{\Vert;\boldsymbol{\hat{n}}_{\text{r}}}^{\left({\text{s}}\right)}\left(r_{\Vert}\right)$ become structureless
 upon heating from the lamellar state to the nematic state. Nevertheless, it should be noted that the 
 transformation is accompanied by considerable hysteresis (see solid/open symbols in Fig.~\ref{fig:03}).
 
Lamellar microphase-separated structures have also been observed in purely entropic mixtures of hard rods and hard spheres \cite{Dogic2000}, as well
as in GB-Lennard Jones mixtures where the strength of rod-sphere attraction is isotropic \citep{Antypov2004}. In both cases, lamellar phases occur already at $\sigma_{s}^{*}=1$. In the present system, 
the low-temperature state at $\sigma_{s}^{*}=1$ is {\em not} lamellar [see Fig.~\ref{fig:02}(c)] but rather smectic-like
with magnetic chains aligned parallel to the smectic director. This satisfies not only the interactions within each species but also the rod-sphere interactions which
favor planar anchoring. Only at $\sigma_{s}^{*}=2$, where the spheres perturb the smectic order sufficiently strongly, the magnetic particles finally
form a layer in their own.  

In conclusion, our MC simulation study of binary mixtures of GB rods and dipolar soft spheres with comparable size ratios (such as in real colloidal-rod systems \citep{May2014,privcomm})
suggests intriguing ordering scenarios of these model LC-MNP hybrid systems not known so far. 
In particular, we have observed (i) spontaneous formation of ferromagnetic chains,
yielding an uniaxially ordered hybrid without net magnetization but enhanced ordering of the LC matrix, and (ii) biaxial nematic and lamellar phases, where the magnetic particles arrange into chains oriented perpendicular to the
LC director and where, at sufficiently low temperatures, local translational ordering resembling a rhombic structure is formed. This novel biaxial LC-MNP system incorporates features of biaxial phases in general \cite{Tschierske2010,Borshch2013} and, in particular, of biaxial mixtures of rod-disk mesogens \cite{Cuetos2008}. Due to the magnetic component, both the uniaxial and biaxial ordered ”hybrids” can be expected to possess strong anisotropic sensitivity to external magnetic and electric fields, as well as to shear forces, with consequences that are yet to be explored. Moreover, the different ordering scenarios will have influence on light propagation and thermal transport in corresponding real soft magnetic hybrids. Our MC results may stimulate further experimental investigations of such systems.

\begin{acknowledgments}
We thank R.~Stannarius and S.~Odenbach for stimulating discussions. Financial support from the German Science Foundation (DFG) via the priority programme SPP 1681 is gratefully acknowledged.
\end{acknowledgments}


%

\end{document}


\noindent
{\bf \Large Supplementary Information for:}\\

\noindent
{\bf \Large Spontaneous ordering of magnetic particles in liquid crystals: From chains
to biaxial lamellae}\\

\noindent
{Stavros D. Peroukidis and Sabine~H.L. Klapp}\\
{\em Institute of theoretical Physics, Secr. EW 7-1, Technical University of Berlin, Hardenbergstr. 36, D-10623 Berlin, Germany}\\
\noindent
{  }\\
\noindent
This supplementary information is organized in two sections: In Sec.~I we present 
definitions for various pair distribution functions which we used to analyze our systems. Numerical results for rod-sphere mixtures 
with size ratios $\sigma_{\text{s}}^{*}=1$ and $\sigma_{\text{s}}^{*}=2$ 
are presented in Sec.~II.

\section{Definitions of distribution functions}
To analyze the positional ordering we consider various direction-dependent pair correlation functions in addition to the usual 
radial pair correlation functions $g^{({\text{a}})}\left(r\right)$ \cite{Veerman1992,McGrother1996,Bebo2000}, where a= r (rods) or s (spheres).
To start with, we calculate the longitudinal correlation function
\begin{equation}
\label{g_par}
g_{\Vert;\boldsymbol{\hat{n}}_{\text{a}}}^{\left({\text{a}}\right)}\left(r_{\Vert}\right)= \Big{\langle} \dfrac{\sum_{i\neq j}\delta\left(r_{\Vert}- \left|\boldsymbol{r}_{ij}\cdot\boldsymbol{\hat{n}}_{\text{a}}\right|  \right)}{\Delta V_2\rho\left(N_{\text{a}}-1\right)}   \Big{\rangle},
\end{equation} 
where $\Delta V_2=\pi\left( r^2 -\left(\boldsymbol{r}_{ij}\cdot\boldsymbol{\hat{n}}_{\text{a}}\right)^2 \right)\Delta r_{\Vert}$ is the volume of a cylindical shell
with thickness $\Delta r_{\Vert}=0.05\sigma$. The function
$g_{\Vert;\boldsymbol{\hat{n}}_{\text{a}}}^{\left({\text{a}}\right)}\left(r_{\Vert}\right)$ measures the distribution of particles of a specific type along the corresponding
director, i.e., $r_{\Vert}$ is the projection of the connection vector $\boldsymbol{r}_{ij}$ onto $\boldsymbol{\hat{n}}_{\text{a}}$. 

Likewise, the distribution perpendicular to the director of one species is measured by the correlation function
\begin{equation}
\label{g_perp}
g_{\bot;\boldsymbol{
 \hat{n}}_{\text{a}}}^{\left({\text{a}}\right)}\left(r_{\bot}\right) = \Big{\langle} \dfrac{\sum_{i\neq j}\delta\left(r_{\bot}- \sqrt{r_{ij}^2-\left(\boldsymbol{r}_{ij}\cdot\boldsymbol{\hat{n}}_{\text{a}} \right)^{2}}\right)}{\Delta V_3\rho\left(N_{\text{a}}-1\right)}   \Big{\rangle},
 \end{equation}
where $r_{\bot}=|\boldsymbol{r}_{ij}-\left(\boldsymbol{r}_{ij}\cdot\boldsymbol{\hat{n}}_{\text{a}} \right)\boldsymbol{\hat{n}}_{\text{a}}|$, 
and $\Delta V_3=\pi L\left( \left(r_{\bot}+\Delta r_{\bot} \right)^2 - r_{\bot}^2 \right) $ with $L$ being the height of the cylindrical shell. 
 
Further, dipole-dipole correlations along the direction perpendicular to ${\hat{n}}_{\text{s}}$
are quantified by 
\begin{equation}
\label{g_pol}
g_{1;\boldsymbol{\hat{n}}_{\text{s}}}^{\left({\text{s}}\right)}\left(r_{\bot}\right) = \dfrac{\Big{\langle}\sum_{i\neq j}\delta\left(r_{\bot}- \sqrt{r_{ij}^2-\left(\boldsymbol{r}_{ij}\cdot\boldsymbol{\hat{n}}_{\text{s}} \right)^{2}}\right)\cos\theta_{ij}   \Big{\rangle}}{\Big{\langle}\sum_{i\neq j}\delta\left(r_{\bot}- \sqrt{r_{ij}^2-\left(\boldsymbol{r}_{ij}\cdot\boldsymbol{\hat{n}}_{\text{s}} \right)^{2}}\right)   \Big{\rangle}}, 
\end{equation}
 where $\cos\theta_{ij}=\boldsymbol{\hat{\mu}}_{i}\cdot\boldsymbol{\hat{\mu}}_{j}$,
 and $ \boldsymbol{\hat{\mu}}_{i}$ is the dipolar unit vector of particle $i$.

Finally, to analyze the structure with respect to the dipole moment of one dipolar particle, we calculate the two-dimensional correlation function
\begin{equation}
\label{g_two}
g^{({\text{s}})}\left(r_{\Vert},r_{\bot}\right)= \Big{\langle} \dfrac{\sum_{i\neq j}\delta\left(r_{\bot}- \sqrt{r_{ij}^2-\left(\boldsymbol{r}_{ij}\cdot\boldsymbol{\hat{\mu}}_{i} \right)^{2}}\right)\delta\left(r_{\Vert}- \left|\boldsymbol{r}_{ij}\cdot\boldsymbol{\hat{\mu}}_{i}\right|  \right) }{\Delta V\rho\left(N_{\text{s}}-1\right)}  \Big{\rangle}, 
\end{equation}
where $\Delta V=\pi\left( \left(r_{\bot}+\Delta r_{\bot} \right)^2 - r_{\bot}^2 \right)\Delta r_{\Vert} $.

\section{Numerical results for correlation functions}
\subsection{Mixtures with $\sigma^{*}_{\text{s}}=1.0$}
\begin{center}
  \begin{figure}
  \includegraphics[scale=1.0,natwidth=1112,natheight=419]{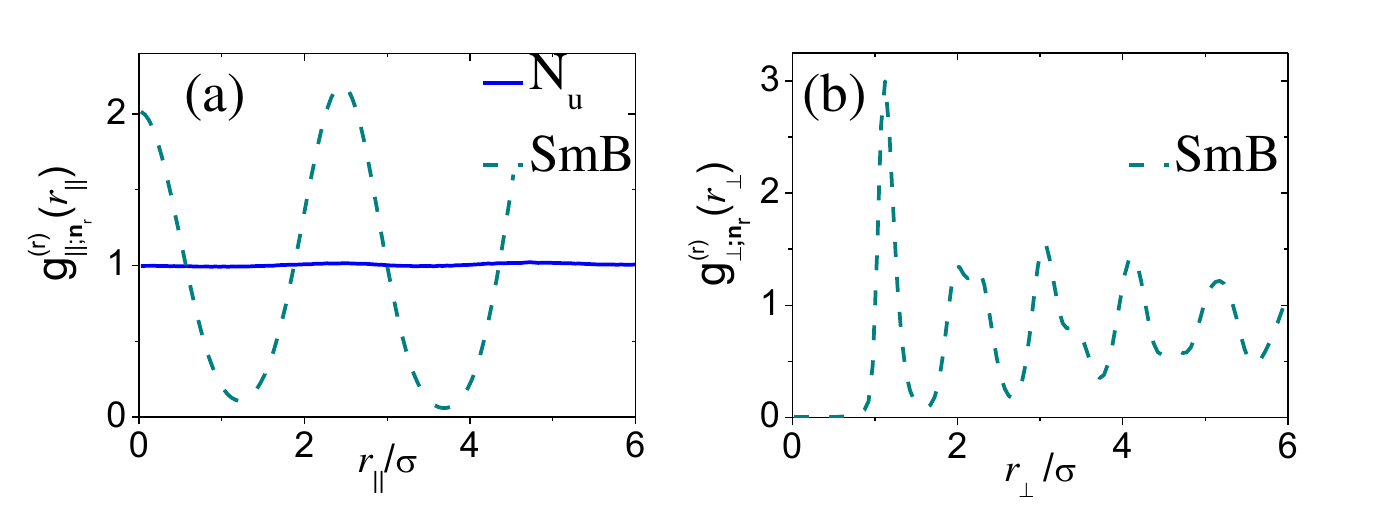}
  \caption{ (Color online) (a) The correlation function 
  $g_{\Vert;\boldsymbol{\hat{n}}_{\text{r}}}^{\left({\text{r}}\right)}\left(r_{\Vert}\right)$ and (b) $g_{\bot;\boldsymbol{\hat{n}}_{\text{r}}}^{\left({\text{r}}\right)}\left(r_{\bot}\right)$ in the uniaxial nematic (N$_{\text{u}}$) state [$(T^*,\rho^{*})=(1.2,0.40)$] and the smectic-B (SmB) state [$(T^*,\rho^{*})=(0.9,0.40)$] at $\sigma^{*}_{\text{s}}=1.0$.}
\label{fig:01_supp}
\end{figure}
\end{center}

We start by considering correlations between the rod particles.
The structure formation of the rods with decreasing temperature at $\sigma_{\text{s}}^{*}=1.0$ (see Figs.~1 and 2 in the main manuscript) is indicated by the correlation function 
$g_{\Vert;\boldsymbol{\hat{n}}_{\text{r}}}^{\left({\text{r}}\right)}\left(r_{\Vert}\right)$ defined in Eq.~(\ref{g_par}).
In Fig.~\ref{fig:01_supp} this function is plotted at density $\rho^{*}=0.4$
for two temperatures pertaining to the nematic and the smectic state, respectively. 
In the uniaxial nematic (N$_{\text{u}}$) state, there are no appreciable correlations. This changes
upon decreasing the temperature, where the undamped oscillations in $g_{\Vert;\boldsymbol{\hat{n}}_{\text{r}}}^{\left({\text{r}}\right)}\left(r_{\Vert}\right)$ clearly reflect the layering typical of a smectic state. The rod correlations $g_{\bot;\boldsymbol{\hat{n}}_{\text{r}}}^{\left({\text{r}}\right)}\left(r_{\bot}\right)$ perpendicular to $ \boldsymbol{\hat{n}}_{\text{r}}$ [see Fig.~\ref{fig:01_supp}(b)] indicate smectic-B (SmB) state characterized by local hexagonal ordering within the layers.

\begin{center}
  \begin{figure}
  \includegraphics[scale=0.80,natwidth=698,natheight=675]{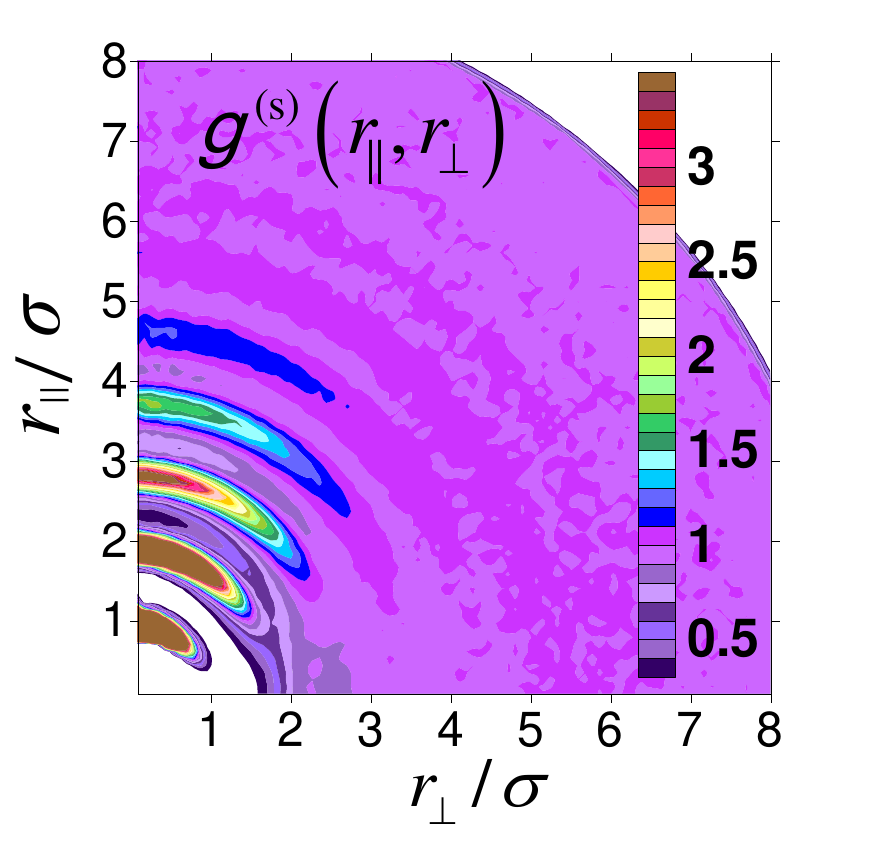}
  \caption{(Color online) Two-dimensional pair correlation function $g^{\left({\text{s}}\right)}\left(r_{\Vert},r_{\bot}\right)$ in the isotropic (I) state at $[(T^*,\rho^{*})=(1.4,0.34)]$ and $\sigma_{\text{s}}^{*}=1.0$.}
  \label{fig:02_supp}
  \end{figure}
\end{center}

We now consider the dipolar particles. In the isotropic (I) state of the LC the dipolar spheres form a (globally isotropic) network of wormlike chains
(see Fig.~2(a) of the main manuscript). A characteristic signature of this arrangement  
are the pronounced correlations along the dipole vector of a given particle. These are visible in Fig.~\ref{fig:02_supp} where we plot the two-dimensional pair correlation function 
defined in Eq.~(\ref{g_two}) at a representative I state.
The correlations along the chains are reflected by the periodic arcs near the $r_{\Vert}$-axis. The lack of correlations
along the $r_{\bot}$-axis indicate that neighboring chains are essentially decoupled, as expected due to the small density of DSS.\\

\begin{center}
  \begin{figure}
   \includegraphics[scale=1.0,natwidth=1113,natheight=417]{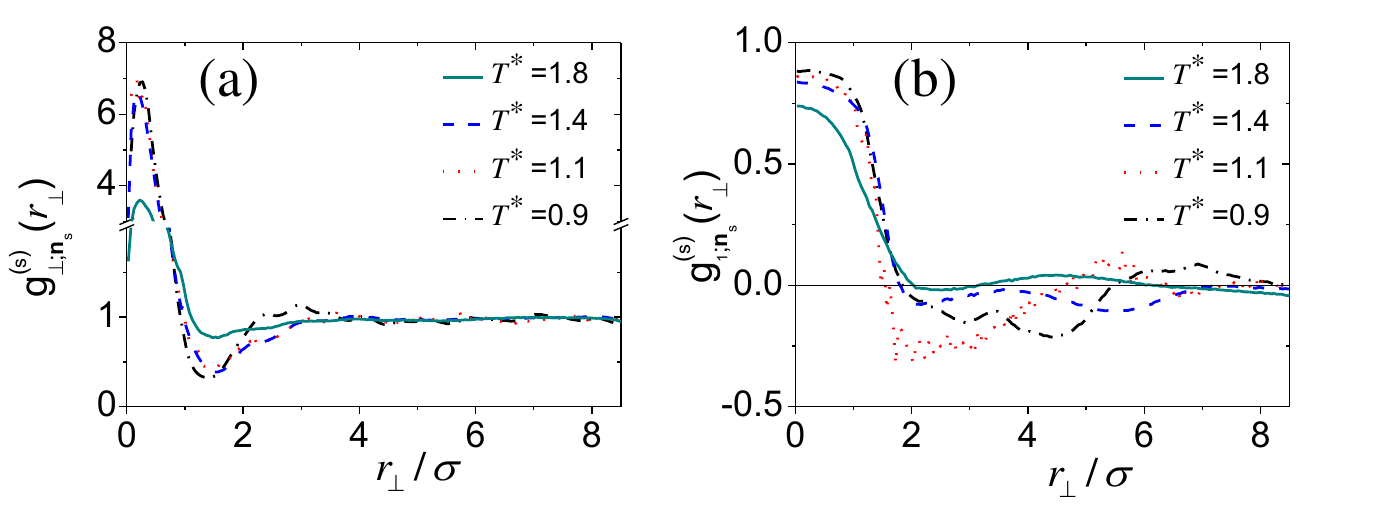} 
  \caption{(Color online) The dipolar correlation functions  (a) $g_{\bot;\boldsymbol{\hat{n}}_{\text{s}}}^{\left({\text{s}}\right)}\left(r_{\bot}\right)$ and
 (b) $g_{1;\boldsymbol{\hat{n}}_{\text{s}}}^{\left({\text{s}}\right)}\left(r_{\bot}\right)$ for $\rho^{*}=0.40$ and various temperatures $T^*$ ranging from the isotropic to the smectic-B regime
 ($\sigma^{*}_{\text{s}}=1.0$).}
  \label{fig:03_supp}
  \end{figure}
\end{center}

Finally, we present in Fig.~\ref{fig:03_supp}
the influence of temperature on correlations between the dipolar spheres (for snapshots, see Fig.~2 in the main manuscript).
Figure~\ref{fig:03_supp}(a) shows that for all temperatures considered, including
the lowest ones corresponding to the uniaxial SmB state, the perpendicular distribution function $g_{\bot;\boldsymbol{\hat{n}}_{\text{s}}}^{\left({\text{s}}\right)}\left(r_{\bot}\right)$
takes large positive values only near the origin and then quickly decays
to one (i.e., no correlations). Similarly, the dipole-dipole correlation function $g_{1;\boldsymbol{\hat{n}}_{\text{s}}}^{\left({\text{s}}\right)}\left(r_{\bot}\right)$,
which is sensitive to the particle orientations, is positive (parallel orientation) only at very small
distance in perpendicular direction. Both functions thus indicate that there are no polar domains beyond
a single chain. Moreover, the polar chains lack of any long-range translational
order.
\subsection{Mixtures with $\sigma^{*}_{\text{s}}=2.0$}
\begin{center}
  \begin{figure}
\includegraphics[scale=1.0,natwidth=1182,natheight=434]{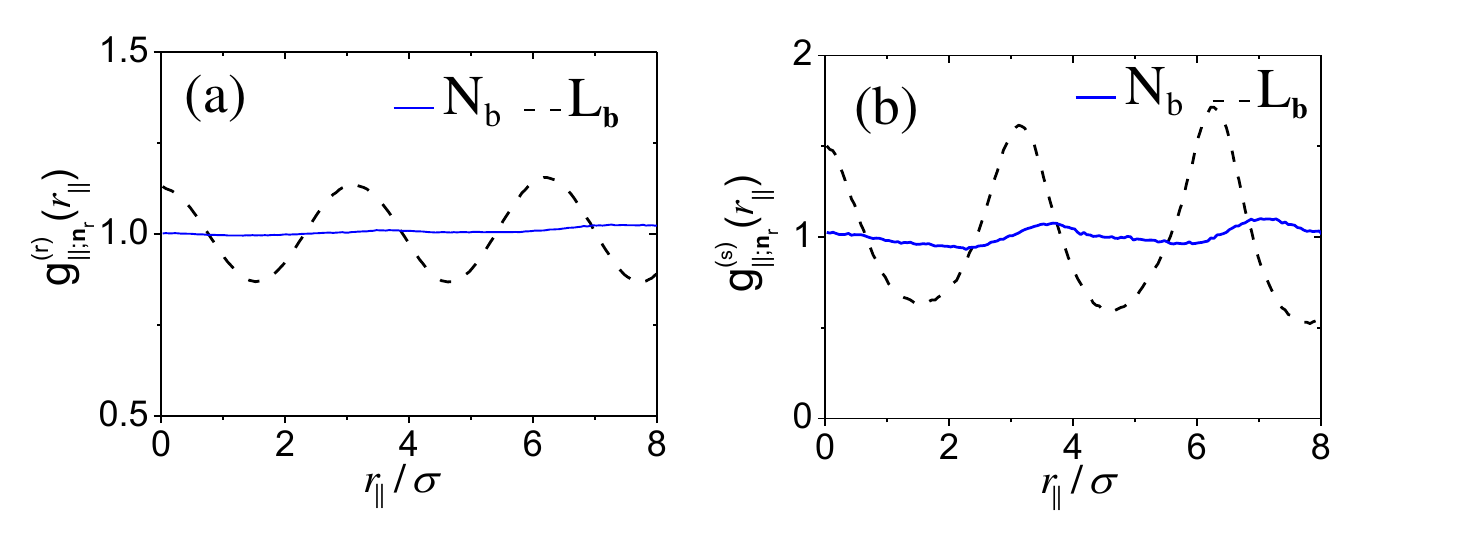}
  \caption{(Color online) The distribution functions (a) $g_{\Vert;\boldsymbol{\hat{n}}_{\text{r}}}^{\left({\text{r}}\right)}\left(r_{\Vert}\right)$ and (b) $g_{\Vert;\boldsymbol{\hat{n}}_{\text{r}}}^{\left({\text{s}}\right)}\left(r_{\Vert}\right)$ along the nematic director in the biaxial nematic (N$_{\text{b}}$) state at $[(T^*,\rho^{*})=(1.8,0.338)]$ and the biaxial lamellar (L$_{\text{b}}$) state at $[(T^*,\rho^{*})=(1.2,0.338)]$. The size ratio
  is $\sigma^{*}_{\text{s}}=2.0$. The data pertain to the system size $N=2536$.}
  \label{fig:04_supp}
  \end{figure}
\end{center}

The mixtures with  $\sigma^{*}_{\text{s}}=2.0$ are characterized by the appearance of a biaxial nematic (N$_{\text{b}}$) and, at even lower temperature, a biaxial lamellar (L$_{\text{b}}$) state
(see Figs.~3 and 4 in the main manuscript).
To understand the differences between these states on the level of correlation functions, we
plot in Fig.~\ref{fig:04_supp} exemplary results for the rod-rod distribution function $g_{\Vert;\boldsymbol{\hat{n}}_{\text{r}}}^{\left({\text{r}}\right)}\left(r_{\Vert}\right)$ 
and the sphere-sphere distribution function  $g_{\Vert;\boldsymbol{\hat{n}}_{\text{r}}}^{\left({\text{s}}\right)}\left(r_{\Vert}\right)$.
At $T^{*}=1.8$ (N$_{\text{b}}$ state), both distribution functions are essentially structureless, indicating that neither the rods nor the spheres are correlated along the nematic director. Upon further cooling
into the L$_{\text{b}}$ state, the function $g_{\Vert;\boldsymbol{\hat{n}}_{\text{r}}}^{\left({\text{r}}\right)}\left(r_{\Vert}\right)$ [see Fig.~\ref{fig:04_supp}(a)] develops oscillations characteristic of the
formation of a layered structure. Interestingly, this is also reflected by oscillations of the
sphere distribution function $g_{\Vert;\boldsymbol{\hat{n}}_{\text{r}}}^{\left({\text{s}}\right)}\left(r_{\Vert}\right)$ along the {\em rod}  $\boldsymbol{\hat{n}}_{\text{r}} $ director [see Fig.~\ref{fig:04_supp}(b)]; these oscillations reflect the formation of dipolar layers between the rods (see Fig.~4(b) in the main manuscript). Indeed, the large 
peak at $r_{\Vert}\approx 3.13\sigma$ occurs at nearly the same value as that in $g_{\Vert;\boldsymbol{\hat{n}}_{\text{r}}}^{\left({\text{r}}\right)}\left(r_{\Vert}\right)$, reflecting
that the length of the rods is the dominant length scale here.

\begin{center}
  \begin{figure}
  \includegraphics[scale=1.0,natwidth=1142,natheight=449]{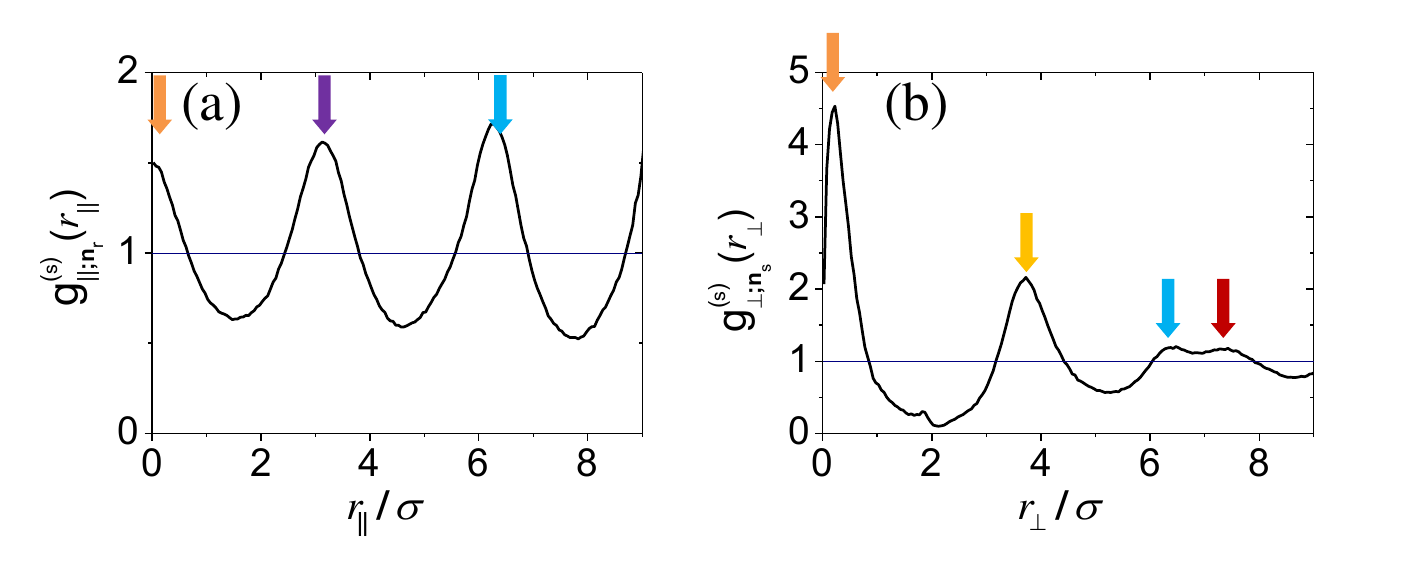}
  \caption{(Color online) The distribution functions (a) $g_{\Vert;\boldsymbol{\hat{n}}_{\text{r}}}^{\left({\text{s}}\right)}\left(r_{\Vert}\right)$ and (b) $g_{\bot;\boldsymbol{\hat{n}}_{\text{s}}}^{\left({\text{s}}\right)}\left(r_{\bot}\right)$ in the biaxial lamellar (L$_{\text{b}}$) state [$(T^*,\rho^{*})=(1.2,0.338)$, $\sigma^{*}_{\text{s}}=2.0$]. 
 The maxima of the correlation function  that are depicted by arrows correspond to distances 
 in the sketch of the chain arrangement in Fig.~\ref{fig:06_supp}.}
  \label{fig:05_supp}
  \end{figure}
\end{center}

Additional information about the distribution of spheres is gained by the function
$g_{\bot;\boldsymbol{\hat{n}}_{\text{s}}}^{\left({\text{s}}\right)}\left(r_{\bot}\right)$
which is sensitive to structure formation perpendicular to the director characterizing the magnetic chains. Note that this function includes contributions from both,
neighboring chains {\em within} a dipolar layer (see
snapshot in Fig.~4(b)(inset) in the main manuscript) and neighboring chains in adjacent layers. A plot is shown in Fig.~\ref{fig:05_supp}(b), while Fig.~\ref{fig:05_supp}(a) contains
for comparison the function $g_{\Vert;\boldsymbol{\hat{n}}_{\text{r}}}^{\left({\text{s}}\right)}\left(r_{\Vert}\right)$ [same data as in Fig.~\ref{fig:04_supp}(b)].

\begin{center}
  \begin{figure}
  \includegraphics[scale=1.0,natwidth=449,natheight=489]{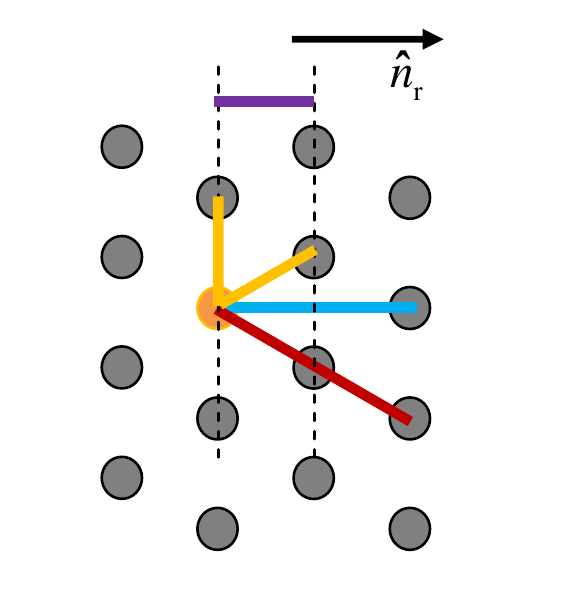}
  \caption{(Color online) Sketch of the (idealized) arrangement of magnetic chains inside the L$_{\text{b}}$ state (for snapshots, see Fig.~4(b) main manuscript). The chains point randomly into or out of the plane. Also indicated is the director of the rods.}
  \label{fig:06_supp}
  \end{figure}
\end{center}

It is seen that $g_{\bot;\boldsymbol{\hat{n}}_{\text{s}}}^{\left({\text{s}}\right)}\left(r_{\bot}\right)$ [as does $g_{\Vert;\boldsymbol{\hat{n}}_{\text{r}}}^{\left({\text{s}}\right)}\left(r_{\Vert}\right)$]
displays pronounced maxima indicative
of the presence of short-ranged translational order of the ferromagnetic chains. Analyzing the positions of the maxima of both functions (indicated by colored arrows) we can draw a sketch
of the arrangement of chains in the L$_{\text{b}}$ state, see Fig.~\ref{fig:06_supp}. Here, the chains point into (or out of) the plane, whereas the rod's director lies horizontally. Thus, the chains
are positioned within layers squeezed between the rods (see snapshot in Fig.~4(b) in the main manuscript).
Of course, the sketch illustrates a highly idealized situation disregarding thermal fluctuations.

The spacing between the dipolar layers can be inferred from the positions of maxima in $g_{\Vert;\boldsymbol{\hat{n}}_{\text{r}}}^{\left({\text{s}}\right)}\left(r_{\Vert}\right)$ (see violet and
light blue colors). The second peak of $g_{\bot;\boldsymbol{\hat{n}}_{\text{s}}}^{\left({\text{s}}\right)}\left(r_{\bot}\right)$ stems from chains in the adjacent layer, on the one hand, and
from the two neighboring chains inside the layer, on the other hand.  Interestingly, this peak (yellow) is positioned at a somewhat larger distance ($r_{\bot}\approx 3.7\sigma$) than that in
the longitudinal function ($r_{\Vert}\approx 3.13\sigma$). Thus, the chains in adjacent layers are shifted, as indicated in Fig.~\ref{fig:06_supp}. Finally, the light blue and
red peaks in $g_{\bot;\boldsymbol{\hat{n}}_{\text{s}}}^{\left({\text{s}}\right)}\left(r_{\bot}\right)$ stem from correlations with chains the next nearest neighboring layer. Altogether,
the correlations reflect that the chains are locally arranged into a rhombic-like structure. 

\bibliography{MyReferencesSupplem.bib}